\documentclass{JHEP3}
\usepackage{amsmath}
\usepackage{amssymb}
\usepackage{epsfig}
\usepackage{latexsym}

\newcommand{\B}{{\cal B}}
\newcommand{\al}{{\alpha}}
\newcommand{\m}{{\text{m}}}
\newcommand{\g}{{\text{gh}}}

\newcommand{\N}{{\cal N}}
\newcommand{\bpz} {{\text{\boldmath{$\flat$}}}}

\newcommand{\her} {\dag}
\newcommand{\Om}{{\Omega}}
\newcommand{\la}{{\lambda}}
\newcommand{\e}{{\epsilon}}

\newcommand{\HH}{{\cal H}}

\newcommand{\ka}{{\kappa}}

\newcommand{\One} {{\bf 1}} 
\newcommand{\tr} {\operatorname{tr}}

\newcommand{\bra}[1] {\left<#1\right|}
\newcommand{\ket}[1] {\left|#1\right>}
\newcommand{\braket}[2] {\left<\vphantom{#2}{#1}\right|
                         \left.\!\vphantom{#1}{#2}\right>}
\newcommand{\braketi}[3] {{\vphantom{\braket{#2}{#3}}}_{#1}
                            \!\braket{#2}{#3}}
\newcommand{\brai}[2] {{\vphantom{\bra{#2}}}_{#1}\!\bra{#2}}
\newcommand{\SL} {\operatorname{SL}}

\newcommand{\va}{\downarrow}
\newcommand{\vb}{\uparrow}

\title{Normalization anomalies in level truncation calculations}

\author{Ehud Fuchs\\
Max Planck Insitut f\"ur Gravitationsphysik\\
Albert Einstein Institut\\
14476 Golm, Germany\\
\email{udif@aei.mpg.de}
}
\author{Michael Kroyter\\
School of Physics and Astronomy\\
The Raymond and Beverly Sackler Faculty of Exact Sciences\\
Tel Aviv University, Ramat Aviv, 69978, Israel\\
\email{mikroyt@tau.ac.il}
}

\abstract{
We test oscillator level truncation
regularization in string field theory
by calculating descent relations among vertices,
or equivalently, the overlap of wedge states.
We repeat the calculation using bosonic,
as well as fermionic ghosts,
where in the bosonic case we do the calculation both
in the discrete and in the continuous
basis. We also calculate analogous expressions in
field level truncation.
Each calculation gives a different result.
We point out to the source of these differences and in the
bosonic ghost case we pinpoint the origin of the difference
between the discrete and continuous basis calculations.
The conclusion is that level truncation regularization
cannot be trusted in calculations involving normalization
of singular states, such as wedge states,
rank-one squeezed state projectors
and string vertices.
}

\preprint{TAUP-2811-05\\AEI-2005-129\\{\tt hep-th/0508010}}

\begin{document}

\section{Introduction}

The number of fields in string field theory~\cite{Witten:1986cc}
is infinite.
In string theory the infinite number of fields
is of great importance
for the natural renormalizability of the theory.
In string field theory the infinite number of fields is a source for
both conceptual and technical challenges.
Level truncation offers a practical solution for the technical
problems, but has no bearing on the conceptual side,
for the simple reason that there is no proof that level truncation is
a consistent regularization.

There are several different level truncation schemes.
In field level truncation~\cite{Kostelecky:1990nt} the level refers to
the field's total level, and fields above a certain level are ignored.
In oscillator level truncation, oscillators
up to a certain level are used and {\it all} possible fields generated
by these operators are considered.

Field level truncation was much used in the study of tachyon
condensation. The tachyon potential should be background
independent~\cite{Sen:1999xm}. Thus, it is enough to focus on the
universal subspace generated by the matter Virasoro generators and by
the $b,c$ ghosts~\cite{Sen:1999nx,Rastelli:2000iu}.
The (physical) universal subspace of ghost number one can be
generated by matter and ghost Virasoro generators.
In this context the tachyon vacuum was evaluated to a remarkable
accuracy~\cite{Gaiotto:2002wy}.
Field level truncation was also used to study other problems, such as
spontenuos Lorentz and CPT breaking~\cite{Kostelecky:1995qk}.

The calculation of D-brane tension ratio in vacuum string field theory
is an example for the use of oscillator level
truncation~\cite{Rastelli:2001jb}.
Taylor suggested using oscillator level truncation
as a way to calculate perturbative diagrams in
string field theory~\cite{Taylor:2002bq}.
Oscillator level truncation
is the main subject of this paper and is what we mean by
level truncation unless stated otherwise.
We also address
field level truncation in the universal basis,
though to a more limited extent.

The use of level truncation may be problematic in light of the fact that
it does not respect the Virasoro symmetry, especially in calculations
that involve infinities that should be
somehow regularized.
In field theory the primary test for a
regularization scheme is that it
respects the symmetries involved.
It was demonstrated in various contexts that similar considerations
should hold in string field
theory~\cite{Potting:1988ra,Okuyama:2002tw,Bars:2002bt}.

In order to avoid the potential problems of
level truncation, one can try to use the method of diagonalizing
the infinite-size matrices involved in the definition of the
star product.
The zero-momentum matter vertex was diagonalized
in~\cite{Rastelli:2001hh}. Following this work, the vertex including
the zero mode and the ghost sector was diagonalized
in~\cite{Feng:2002rm,Belov:2002fp,Erler:2002nr,Belov:2002pd}.
The diagonal form of the vertex
drastically simplifies calculations,
at least for some star-subalgebras~\cite{Fuchs:2002zz,Ihl:2003fw}.
It also carries the promise of getting analytical results in
string field theory~\cite{Okuyama:2002yr,Okuyama:2002tw}.
However, an anomaly
in descent relations and wedge state inner products in the diagonal
basis was found in~\cite{Belov:2002pd,Fuchs:2002wk,Belov:2002sq}.
An interesting interpretation of this anomaly was
given in~\cite{Belov:2003qt}.
Here, we re-examine this anomaly and relate it to an
anomaly in the level truncation calculation.

Our testing ground for level truncation is the
$\braket{V_3}{V_1}$ descent relation.
String vertices describe string interactions by gluing, with the
$N$-vertex $\bra{V_N}$ gluing $N$ string fields, according to,
\begin{equation}
\label{VnDef}
\brai{N..1}{V_N}\ket{\Psi_1}_1..\ket{\Psi_N}_N=
  \int \Psi_1\star..\star\Psi_N\,,
\end{equation}
where the star product and integral implement the
gluing~\cite{Witten:1986cc}.
Since string vertices
can be considered as states in multi-string spaces, they can be
glued by other vertices. This fact induces the following descent relations
among the string vertices,
\begin{equation}
\label{VnIdentities}
\brai{1..n}{V_n}\brai{(n+1)..(n+m)}{V_m}\ket{V_2}_{1,(n+m)}=
\brai{2..(n+m-1)}{V_{n+m-2}}\,.
\end{equation}
The normalization constant of this relation for $n=3,m=1$ is what was
found to be anomalous when evaluated using bosonized ghosts in the
continuous basis.
It was not clear whether the source of the anomaly
is the level truncation, the use of bosonic ghosts or the
use of the continuous basis.
The relation of the continuous basis to level truncation lies in the
fact that divergent quantities are regularized
by relating them to $\log \ell$, where $\ell$ is the
truncation level.

In this paper, we repeat the calculation in the discrete
basis for both bosonic and fermionic ghosts.
The results we get are different
from the previous calculations, but still anomalous.
For the bosonic ghost, we discuss these differences in
section~\ref{sec:bosonic} and show explicitly how the calculation in
the discrete basis can be modified to give the result of the
continuous basis.
In section~\ref{sec:fermi} we perform the fermionic ghost calculation
and get a new value for the anomaly.
We believe that the reason for the discrepancy between the
bosonic and fermionic ghost calculations
is the non-linear
relation between
them, $b,c=:e^{\pm\rho}:$.

In light of the success of level truncation calculations,
in particular those of~\cite{Taylor:2002bq}, our claim
of an anomaly needs an explanation.
In~\cite{Taylor:2002bq}, Taylor had to integrate over a parameter
$T$, representing the length of an intermediate string.
The limit $T\rightarrow0$ corresponds to calculating the descent relation
$\bra{V_3}\bra{V_3}\ket{V_2}=\bra{V_4}$.
This is a singular limit of the integrands in~\cite{Taylor:2002bq}, but
the singularity is integrable.
The calculation at $T=0$ is anomalous, but it does not contribute to
the integral.

Another discrepancy with~\cite{Taylor:2002bq} lies in the form of the
fits. For arbitrary expression $X$, we use fits of the form
$X=a+b/\ell^c$, where $\ell$ is the level and $a,b,c$ are parameters,
while in~\cite{Taylor:2002bq}, $c=1$ always gives good fits.
The fact that our values for $c$ are not too far from unity (they are
in the range 0.2 .. 2.0), together with the observation
in~\cite{Taylor:2002bq} that convergence becomes slower as $T=0$ is
approached, resolve this issue.
The value of $c$ was studied in a related context
in~\cite{Beccaria:2003ak}.

An alternative description of the anomaly is in terms of inner product
of specific wedge states~\cite{Rastelli:2000iu}.
Wedge states are surface states~\cite{LeClair:1989sp,LeClair:1989sj},
and as such are described as
exponentials of Virasoro generators over particular vacua.
Wedge states are also easy to deal with since they are diagonal in the
continuous basis~\cite{Rastelli:2001hh}.
However, the continuous basis representation can be regarded as
a generic source of singularity for diagonal
states, since their defining matrix is proportional to
$\delta(\ka-\ka')$.
This is the origin of anomalies in this case.

Field level truncation is natural for describing
surface states and wedge states in particular.
In section~\ref{fieldLT}, we suggest that anomalies are possible in
this description as well, despite the fact that we cannot trace them
in the level truncation using
Virasoro generators with vanishing central carge, $c=0$.
To that end we use field level truncation with Virasoro
generators of arbitrary $c$.
We also compare these calculations to the analogous ones in the matter
sector using oscillator level truncation and continuous basis
techniques.
We conclude in section~\ref{sec:conc} by summarizing our results and
suggesting future research directions.

\section{Bosonic ghost}
\label{sec:bosonic}

In this section we
calculate the descent relations using the bosonic ghost.
We start by setting our conventions in~\ref{BosVert},
then we evaluate the normalization in~\ref{BosAnom}.
There is no ``simple'' way to calculate the normalization.
We perform the calculation in the discrete and in the continuous
basis using level truncation.
In the continuous basis, level truncation
manifests itself by regulating infinities using expressions that are
proportional to $\log \ell$.
In both calculations the ``infinite''
contributions cancel each other, but the final result is anomalous,
with different results in each calculation.
We end by comparing the anomalies in these two cases.

\subsection{The vertices}
\label{BosVert}

In the matter sector the form of the vertices is,
\begin{equation}
\ket{V_N(P_0)} = \delta
\Big(\sum_{i=1}^N p_0^i\Big)
\exp\left(
    -\frac{1}{2}A^\dagger V_N A^\dagger
    +P_0 V_{N0} A^\dagger
    -\frac{1}{2}P_0 V_{N00} P_0 \right)
\ket{0}_N\,.
\end{equation}
$P_0$ is an $N$-vector of the zero-momenta $p_0^i$.
$A^\dagger$ is an $N$-vector of creation operators in the $N$ Fock
spaces, which also has a hidden index $n$ for the excitation number
inside the Fock space.
$V_N$ is an $N$ by $N$ matrix of infinite bi-linear forms.
$V_{N0}$ is an $N$ vector of infinite vectors,
$V_{N00}$ is an $N$ by $N$ matrix of scalars and $\ket{0}_N$
is the direct product of $N$ Fock space vacua.
To keep things simple, we are suppressing the 26 dimensional space-time
indices, but it is crucial to have 26 dimensions for the cancellation
of the infinite normalization.

For the identity state ($N=1$),
$V_1$ is the twist matrix $C$, $V_{10}=0$ and $V_{100}=0$.
For the two-vertex
$V_2^{12}=V_2^{21}=C$,
$V_2^{11}=V_2^{22}=0$, $V_{20}=0$ and $V_{200}=0$.
The three-vertex is a bit more complicated. Its coefficients in the
regular oscillator basis are summed up nicely
in~\cite{Rastelli:2001jb}. For the continuous basis, they can be found
in~\cite{Fuchs:2003wu}.

The vertex in the bosonic ghost sector is similar, but with some
modifications due to the anomalous ghost current,
\begin{equation}
\ket{V^g_N(Q_0)} = \delta_{\sum_{i=1}^N q_0^i-\frac{3}{2}(N-2)}\exp\left(
    -\frac{1}{2}A^\dagger V^g_N A^\dagger
    +Q_0 V^g_{N0} A^\dagger
    -\frac{1}{2}Q_0 V^g_{N00} Q_0 \right)
\ket{0}_N.
\end{equation}
$Q_0$ is an $N$-vector of zero-momenta $q_0^i$, which are discrete
half-integer numbers. That is why the Dirac delta becomes
a Kronecker delta.
To describe the coefficients in the ghost sector, it is useful to
introduce the vector $J$ that relates the mid-point to the
zero mode,
\begin{align}
X\big(\frac{\pi}{2}\big) &=
x_0+\sum_{n=1}^\infty J_n x_n = x_0 + \int d\ka J_\ka x^\ka\,,\\
J_{2n} &= \sqrt{2}(-1)^n\,,\quad J_{2n+1} = 0\,,\\
J_\ka &= {\cal P}\frac{\sqrt{2}}{\ka\sqrt{{\N}(\ka)}} \,,
\end{align}
where as usual,
\begin{equation}
\N(\ka)\equiv \frac{2}{\pi}\sinh\big(\frac{\ka \pi}{2}\big)\,.
\end{equation}

For the identity state, $V^g_1=V_1=C$, $V^g_{10}=\sqrt{2}J$ and
$V^g_{100}=0$. For the two-vertex, the bosonic ghost coefficients are
identical to the matter sector. This is a result of the fact that the
constant term in the Kronecker delta is zero in this case.
For the three-vertex $V^g_3=V_3$,
$V^g_{30}=V_{30}+\frac{\sqrt{2}}{3}J$ and
$V^g_{300}=V_{300}-\frac{3}{2}\log\left(\frac{27}{16}\right)$.
One way to calculate the last constant is from the
identity,
\begin{equation}
\bra{q_0^1=-\frac{3}{2}}
\bra{q_0^2=-\frac{3}{2}}
\bra{q_0^3=\frac{3}{2}}\ket{\vphantom{\frac{3}{2}}V_3}=1\,.
\end{equation}

\subsection{The anomaly}
\label{BosAnom}

We can now use the results of~\cite{Kostelecky:2000hz} to calculate,
\begin{equation}
\braket{V_1}{V_3} = \gamma_{13} \ket{V_2}\,.
\label{DescRel}
\end{equation}
This is the first non-trivial descent relation.
If we use a consistent regularization, we should get $\gamma_{13}=1$.
The expression for $\gamma_{13}$ is,
\begin{equation}
\begin{aligned}
\label{ga13}
\gamma_{13}=\det(\One-C V_3^{11})^{-\frac{D+1}{2}}
\exp\frac{9}{4}\Bigl(&
-\frac{1}{2}V_{30}^{g11}(\One-C V_3^{11})^{-1} C
V_{30}^{g11}
+V_{10}^g(\One-V_3^{11} C)^{-1}
V_{30}^{g11}\\
&-\frac{1}{2}V_{10}^g(\One-V_3^{11} C)^{-1} V_3^{11} V_{10}^g
-\frac{1}{2}V_{300}^{g11}\Bigr).
\end{aligned}
\end{equation}
This computation involves infinite matrices, therefore
it should be consistently regularized.
The two methods that we can try to deploy are
level truncation in the discrete basis and regularization in the
continuous basis.
The two methods give different result and both results are wrong.
We explain the difference between the two methods and show how the
truncation in the discrete basis can be modified to give the same result
as in the continuous basis.

In level truncation we did the calculation up to level 1000, giving
$\gamma_{13}\approx 0.340$.
This result is summarized in figure~\ref{fig:BosNorm}.
\FIGURE{
\centerline{ \epsfig{figure=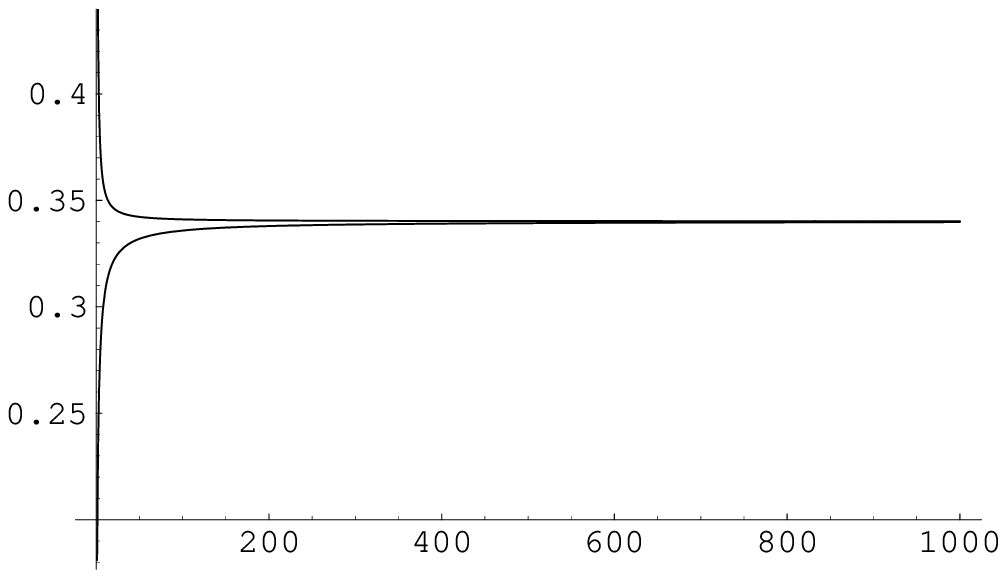, width=10cm} }
\caption{This plot shows the value of $\gamma_{13}$ calculated for levels
  from 1 to 1000. We get different behaviors for even and odd
  levels.
  This is in accord with the difference between odd and even levels
  described in~\cite{Rastelli:2001hh}.
  Both cases fit nicely to a function of the form $a+b / \ell^c$,
  with a common limit value
  $a=0.340$, with $b=0.0948,-0.159$ and $c=1.19,0.91$
  for the even and odd cases respectively.
  Both the calculated points and the fitted curves are plotted in
  this graph, but it is impossible to tell them apart. One remarkable
  feature of the fit is that it fits well starting from level 1.}
\label{fig:BosNorm}
}
The continuous basis calculation was performed
in~\cite{Fuchs:2003wu}. The strategy there was to use the identity,
\begin{equation}
\label{detExpTr}
\det(M) = \exp\big(\tr(\log\, M)\big)\,,
\end{equation}
in order to represent the determinant in a similar way to the other
part of the expression, which can also be represented using a trace.
Then, in order to evaluate the trace of a matrix, which is
diagonal in the continuous basis, the following identity was
used,
\begin{equation}
\tr(A)=\int_{-\infty}^\infty d\ka\,\rho(\ka)a_\ka\,.
\end{equation}
Here, $A$ is such a matrix, $a_\ka$ is the $\ka$ eigenvalue of $A$
and $\rho(\ka)$ is the divergent spectral density.
In level truncation the spectral density is given by,
\vskip 2cm
\begin{equation}
\label{specDens}
\rho^\ell(\ka)=\frac{\log \ell}{2\pi}+\rho^\ell_{\text{fin}}(\ka)\,,
\end{equation}
and $\rho^\ell_{\text{fin}}(\ka)$ has a finite limit for
$\ell \rightarrow \infty$~\cite{Belov:2002pd,Fuchs:2002wk,Belov:2002sq},
\begin{equation}
\rho_{\text{fin}}(\ka)=
\frac{4\log(2)-2\gamma-
       \psi(1+\frac{i \ka}{2})-\psi(1-\frac{i \ka}{2})}{4\pi}\,.
\end{equation}
Here, $\gamma$ is Euler's constant and $\psi$ is the digamma function.
In the evaluation of~(\ref{ga13}), the infinite part (the part that is
proportional to $\log \ell$) dropped out in $D=26$ dimensions and
the final result of the calculation was,
\begin{equation}
\gamma_{13}= 3^{3/8} e^{\frac{3}{2} \gamma+36\zeta'(-1)}
\Big(\frac{\Gamma \big(\frac{1}{3}\big)}{\sqrt{\pi}}\Big)^9
 \approx 0.382948\,,
\end{equation}
where $\zeta$ is the Zeta-function.
The same result was later obtained in~\cite{Belov:2003qt},
where it was given a different interpretation.

Both methods give the wrong result, but the results are suspiciously
close. There are two reasons for the difference between the two
results. One difference comes from the
evaluation of the determinant using~(\ref{detExpTr}).
The calculation in the continuous basis is equivalent to truncating
the matrix $\log M$ and then taking the level to infinity, while in
level truncation we truncate the matrix $M$ itself.
Since $(\log M)_\ell\neq\log(M_\ell)$ we are getting different results.
To demonstrate that this is the source of the difference, we repeat
the level truncation calculation, but this time we calculate the
determinant as follows,
\begin{align}
(\det M)_{\ell,m} = \exp \Big(\tr \big(\log (M_{\ell\cdot m})_{\ell}\big)\Big),
\end{align}
where $\ell$ is the level and $m$ stands for a margin.
Here, we evaluate the matrix to level $m\cdot \ell$, calculate
the log of this matrix, truncate the result to level $\ell$, take the
trace and then exponentiate.

Since the trace diverges, we have to add an infinity of the form
$\frac{3}{2}H_{\ell/2}$,
where $H$ is the harmonic number, to the result.
After adding this term, the result for the trace in the $\ka$
basis is $-1.7448$. For regular level truncation we get
-1.8278 at level 100 and for the corrected level truncation we get
$-1.7411$.
These results are summarized in figure~\ref{fig:DetNorm}.
\FIGURE{
\centerline{ \epsfig{figure=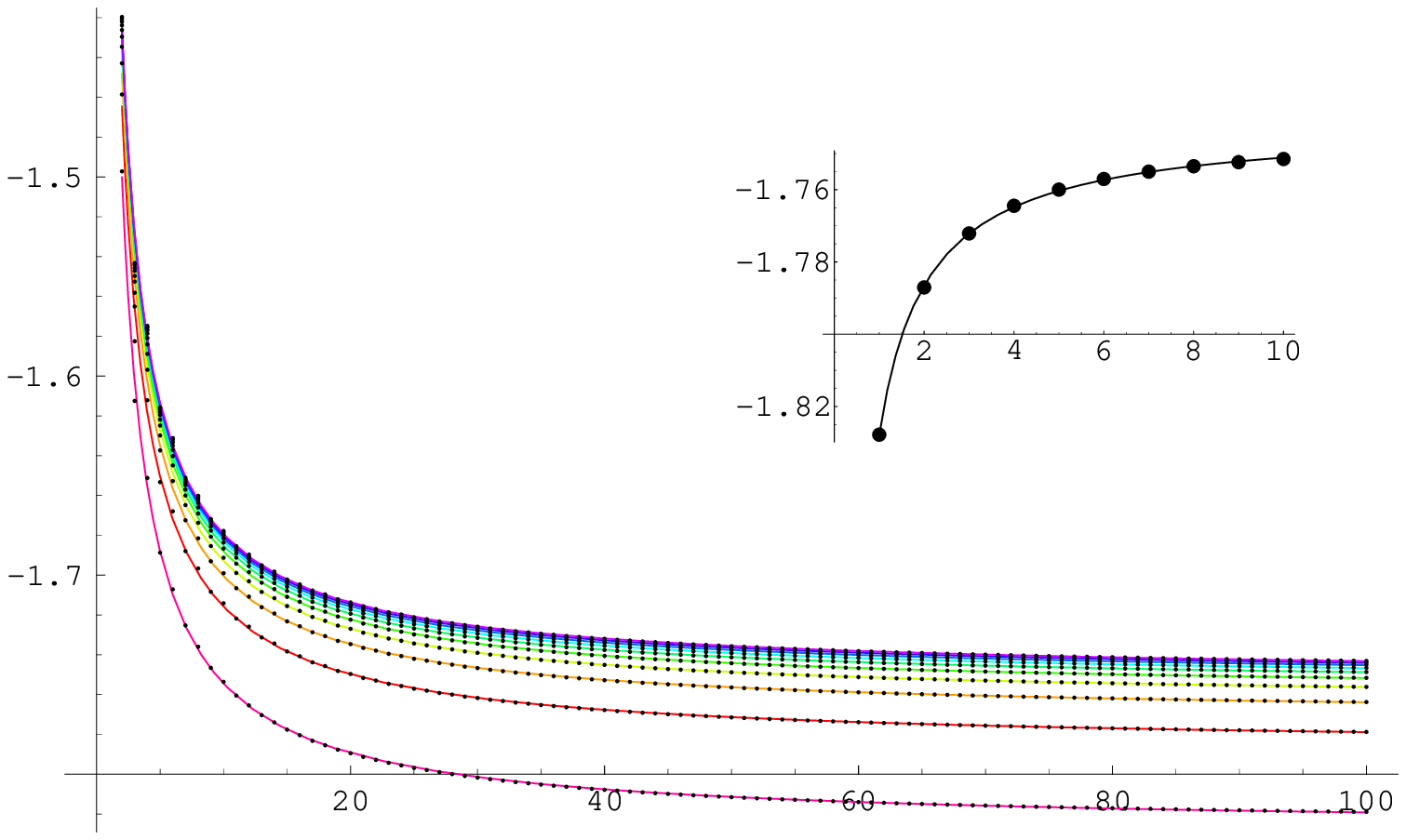, width=5in} }
\caption{This graph shows the effect of the different truncation in the
calculation of the determinant.
The lowest plot in the spectrum has $m=1$, the next one has $m=2$
and so on up to $m=10$. The fit parameters for $a+b/\ell^c$ give
$b$ in the range 0.624 to 0.625,
$c$ in the range 0.93 to 0.94 and
the sub-graph shows the ten values of $a$.
The extrapolation of this graph with the same type of fit
gives us the estimate
of $-1.7411$ (with $b=-0.103, c=1.19$).
If we ignore the first point when extrapolating the sub-graph we will
get a result of $-1.7433$ and ignoring the first five points would
give $-1.7449$. This demonstrates that the
result of $-1.7448$ that we get in the continuous basis
is within the error of the level truncation calculation.
}
\label{fig:DetNorm}
}

The second difference comes from the matrix inversion in the exponent.
Again, the calculation in the continuous basis is equivalent to
truncating the matrix after the inversion, while in level truncation
we first truncate the matrix and only then invert it.
To demonstrate that this is the source of the difference, we repeat
the level truncation calculation but this time we calculate the
matrix to be inverted up to level $m\cdot \ell$, invert the matrix
and only then truncate it to level $\ell$.
This time we have to subtract $\frac{3}{2}H_{\ell/2}$ from the results
in order to overcome the divergences.
After this subtraction, the result in the continuous basis is
$\frac{3}{2}\log\frac{27}{16}\approx 0.7849$.
For the regular level truncation we get $0.7515$ and for
the corrected truncation we get $0.7851$.
These results are summarized in figure~\ref{fig:ExpNorm}.
\FIGURE{
\centerline{ \epsfig{figure=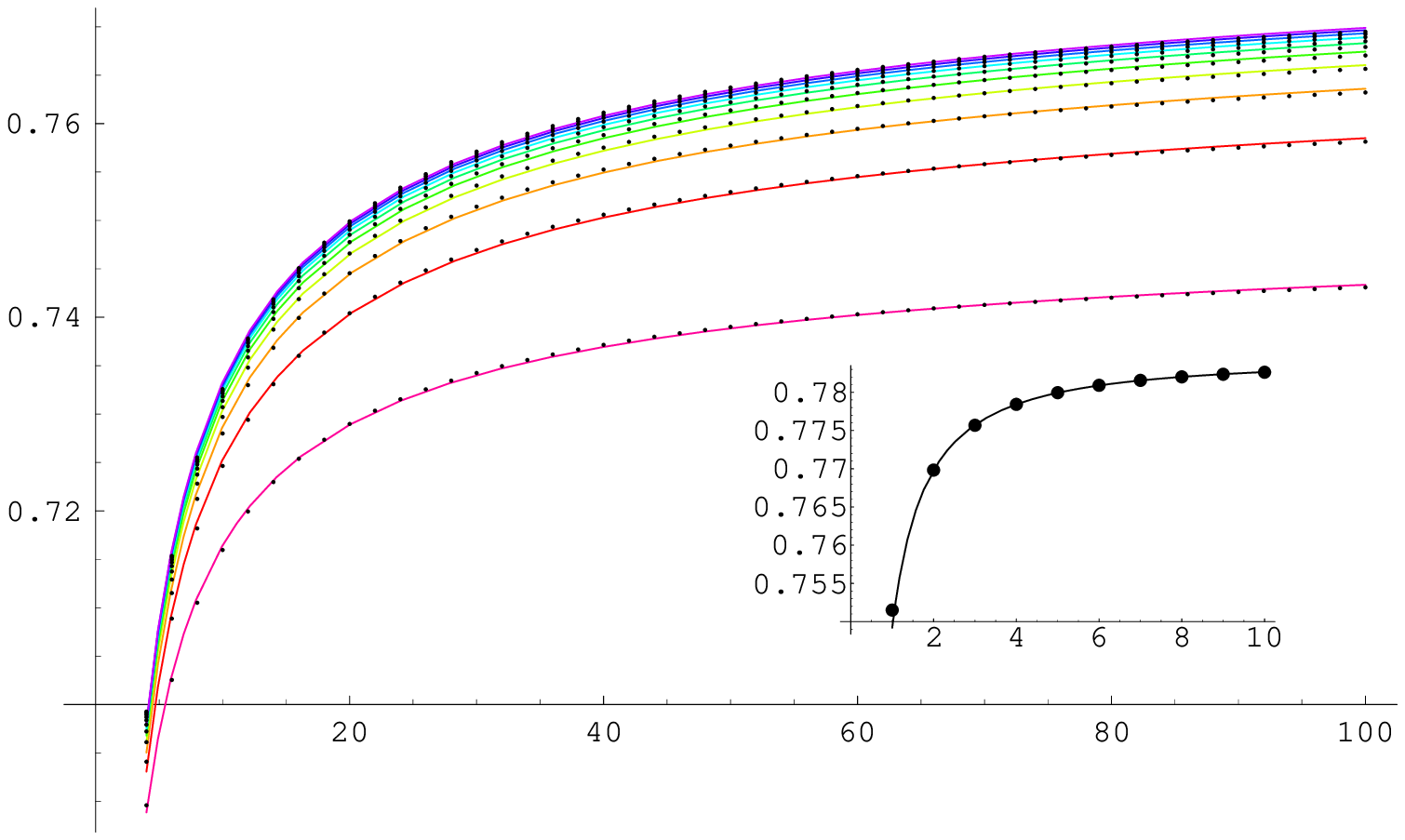, width=5in}}
\caption{This graph shows the effect of the different truncation in the
calculation of the exponent.
The lowest plot in the spectrum has $m=1$, the next one has $m=2$
and so on up to $m=10$.
The fit parameters for $a+b/\ell^c$ give $b$ in the
range -0.15 to -0.19, $c$ in the range 0.63 to 0.59 and
the sub-graph shows the ten values of the extrapolated main plots.
The extrapolation of this graph gives us the estimate of $0.7851$
(with $b=-0.03, c=1.26$).
If we ignore the first point when extrapolating the sub-graph we
will get a result of $0.7846$ and ignoring the first five points
would give $0.7843$. This demonstrates that the
continuous basis result of $0.7849$ is within the error of
the level truncation result.
}
\label{fig:ExpNorm}
}

\section{Fermionic ghost}
\label{sec:fermi}

It was suspected that the
mismatch in the continuous basis calculations
may be the result of using the bosonized ghosts~\cite{Fuchs:2002wk}.
It is clear that level truncation does not commute with
bosonization, which is a non-linear transformation.
In this section, we verify that indeed the result
of the evaluation is changed when using fermionic ghosts in the
level truncation.
However, the result is still not the expected value of unity.
We start by setting our conventions and defining the fermionic ghost
vertices. Then, we turn to the evaluation of the anomaly.
We also confirm that there is no anomaly in the result other
than in the normalization.

\subsection{The vertices}

One of the features of the ghost sector is the existence of several
vacua, due to the ghost zero modes. Here, we follow the conventions
of~\cite{Polchinski:1998rq} by defining,
\begin{equation}
\ket{\va}\equiv c_1 \ket{0} ,
\qquad \ket{\vb}\equiv c_0 c_1 \ket{0}.
\end{equation}
We also define~\cite{LeClair:1989sp},
\begin{equation}
\label{The3Vac}
\ket{3}\equiv c_{-1} c_0 c_1 \ket{0} .
\end{equation}
Each vacuum has its BPZ conjugate,
\begin{equation}
\bra{\va}=\bra{0}c_{-1}\,,\qquad
 \bra{\vb}=\bra{0}c_{-1}c_0\,,
 \qquad \bra{3}=\bra{0}c_{-1}c_0 c_1\,.
\end{equation}
We normalize the vacua in the usual way,
\begin{equation}
\label{norma}
\braket{0}{3}=\braket{\va}{\vb}=\braket{\vb}{\va}=\braket{3}{0}=1\,.
\end{equation}

Vertices of order two and higher are described using tensor products
of string field spaces. In this case our conventions are,
\begin{equation}
\ket{\Psi_1,\Psi_2}_{12}=\ket{\Psi_1}_1 \ket{\Psi_2}_2\,,\qquad
\brai{12}{\Psi_1,\Psi_2}=\brai{1}{\Psi_1}\brai{2}{\Psi_2}.
\end{equation}
Higher multi-string spaces are treated similarly.
For calculations that involve tensor products, we also have to define
the parity of the $\SL(2)$ invariant vacua.
We use the conventions of~\cite{Witten:1986cc}, according to which the
right vacuum $\ket{0}$ is even, while the left vacuum $\bra{0}$ is
odd, though one can also use the opposite
definitions~\cite{Zwiebach:1997fe}.
Both vacua have ghost number zero.

The ghost number of an $N$-vertex is $3(N-1)$.
This follows from the definition of the $N$-vertex~(\ref{VnDef}),
the requirement of having three ghosts to saturate the inner product
in each string field space
and the fact that the string field integral is non-vanishing only
for ghost number three operators.
This is also consistent with the descent relations~(\ref{VnIdentities}).

Taking the parity conventions into account, we deduce that the parity
of $\ket{V_N}$ is $-(-1)^N$, while $\bra{V_N}$ is always odd. This is
also consistent with~(\ref{VnIdentities}).
Our parity conventions imply the following symmetry properties
for the two-string vacua,
\begin{equation}
\begin{aligned}
\label{vacSym}
\ket{0}_{12}&=\ket{0}_{21} ,&\qquad
\ket{\va}_{12}&=-\ket{\va}_{21} ,&\qquad
\ket{\vb}_{12}&=\ket{\vb}_{21} ,&\qquad
\ket{3}_{12}&=-\ket{3}_{21} ,\\
\brai{12}{3}&=\brai{21}{3} ,&\qquad
\brai{12}{\vb}&=-\brai{21}{\vb} ,&\qquad
\brai{12}{\va}&=\brai{21}{\va} ,&\qquad
\brai{12}{0}&=-\brai{21}{0} .
\end{aligned}
\end{equation}
Similar expressions hold for higher tensor products.

The one string vertex $\bra{V_1}$ is the BPZ conjugate of the identity
string field, which is a surface state
defined by the conformal map,
\begin{equation}
\label{fId}
f(z)=\frac{z}{1-z^2}\,.
\end{equation}
The generating function and the defining matrix in the ghost sector
are~\cite{LeClair:1989sp,Maccaferri:2003rz},
\begin{align}
\label{Sghost}
S(z,w)&=\frac{f'(z)^2 f'(w)^{-1}}{f(z)-f(w)}
            \left(\frac{f(w)-f(0)}{f(z)-f(0)}\right)^3
 - \frac{1}{z-w} \left(\frac{w}{z}\right)^3.\\
{(S_0)}_{nm}&=-(-1)^{n+m}\oint_0 \frac{dz dw}{(2\pi i)^2}
   \frac{S(z,w)}{z^{m-1}w^{n+2}}\,.
\end{align}
Note that this differs from the conventions we used
in~\cite{Fuchs:2004xj}, by a sign and by exchanging $n,m$,
due to the different $b,c$ ordering that we use below.
From this expression and the conformal transformation of the
identity~(\ref{fId}) we can read,
\begin{equation}
\begin{aligned}
\ket{V_1}&=\exp\Big(
  \sum_{n=2}^\infty (-1)^{n+1} b_{-n}c_{-n}+
  X_b(c_1-c_{-1})+ 2Y_b c_0
 \Big)\ket{0},\\
X_b&\equiv \sum_{n=1}^\infty (-1)^n b_{-(2n+1)}\,,\qquad
Y_b\equiv \sum_{n=1}^\infty (-1)^n b_{-2n}\,.
\end{aligned}
\end{equation}

In~\cite{Gross:1987fk} a slightly different representation was found
for the one-vertex,
\begin{equation}
\ket{V_1}=N_1
  b_+\big(\frac{\pi}{2}\big) b_-\big(\frac{\pi}{2}\big)
 \exp\Big( \sum_{n=1}^\infty (-1)^{n+1} b_{-n}c_{-n}
 \Big)\ket{\vb}\,.
\end{equation}
Here,
\begin{equation}
b_\pm\big(\frac{\pi}{2}\big)=\sum_{n=-\infty}^\infty
  (\pm i)^n b_n\,,
\end{equation}
and the normalization $N_1$ that was not calculated
in~\cite{Gross:1987fk} can be found from comparing the two
expressions,
\begin{equation}
N_1=\frac{i}{4}\,.
\end{equation}
In the following calculation
we shall find it convenient to use the canonically
conjugate vacua $\bra{\vb}$, $\ket{\va}$.
Thus, we also need a representation of $\ket{V_1}$ over $\ket{\va}$,
\begin{equation}
\label{V1}
\begin{aligned}
\ket{V_1}=&
  \exp\Big(\sum_{n=2}^\infty (-1)^{n+1} b_{-n}c_{-n}+2Y_b c_0
   -X_b c_{-1}\Big)(1+X_b c_1)b_{-1}\ket{\va}=\\
\tilde X_b & \exp\Big(\sum_{n=1}^\infty (-1)^{n+1} b_{-n}c_{-n}+
      2Y_b c_0\Big) \ket{\va},
\end{aligned}
\end{equation}
where we define,
\begin{equation}
\tilde X_b\equiv \sum_{n=0}^\infty (-1)^n b_{-(2n+1)}\,.
\end{equation}

The two-vertex, which implements the BPZ conjugation, is given
by~\cite{LeClair:1989sp},
\begin{equation}
\brai{12}{V_2}=\brai{12}{0}
 (c_1^1-c_{-1}^2)(c_0^1+c_0^2)(c_{-1}^1-c_1^2)
 \exp \!\bigg(\sum_{n=2}^\infty \Big
  (b^{1}_{n} (-1)^n c^{2}_{n}+b^{2}_{n} (-1)^n c^{1}_{n}
 \Big)\bigg)\,.
\end{equation}
Note that the exponent is symmetric with respect to interchanging the
labels ($1,2$), while the zero mode factor is antisymmetric. Since the
vacuum is also antisymmetric~(\ref{vacSym}), we get that the
two-vertex is symmetric. It can also be written as,
\begin{equation}
\label{V2up}
\brai{12}{V_2}=\brai{12}{\vb}(b_0^1-b_0^2)
 \exp \!\bigg(\sum_{n=1}^\infty \Big
  (b^{1}_{n} (-1)^n c^{2}_{n}+b^{2}_{n} (-1)^n c^{1}_{n}
 \Big)\bigg)\,.
\end{equation}

The right two-vertex $\ket{V_2}$ is defined using the relation
(eq. 6.9 of~\cite{LeClair:1989sp}),
\begin{equation}
\label{V2V2def}
\braketi{12}{V_2}{V_2}_{32}=\One_{31}\,,
\end{equation}
which gives,
\begin{equation}
\ket{V_2}_{12}=
 \exp \!\bigg(\!-\sum_{n=2}^\infty \Big
  (b^{1}_{-n} (-1)^n c^{2}_{-n}+b^{2}_{-n} (-1)^n c^{1}_{-n}
 \Big)\bigg)
 (c_1^1-c_{-1}^2)(c_0^1+c_0^2)(c_{-1}^1-c_1^2)\ket{0}_{12}.
\end{equation}
This is the BPZ conjugate of $\bra{V_2}$, provided we identify the
$\ket{0}_{12}$ as the BPZ conjugate of $\brai{12}{0}$.
Note, however, that this identification has a sign ambiguity, since
$\ket{0}_{12}$ is even, while $\brai{12}{0}$ is odd~(\ref{vacSym}).
Also note that $\ket{V_2}$ does not implement BPZ
conjugation\footnote{We do not distinguish here BPZ from
BPZ$^{-1}$.
Related discussion on string vertices can be found
in~\cite{Bonora:2003xp,Kling:2003sb,Maccaferri:2003rz}.}.
To be precise, it implements BPZ conjugation up to
a sign, as implied by~(\ref{V2V2def}),
\begin{equation}
\ket{\Psi}_3=\braketi{12}{V_2}{V_2}_{32}\ket{\Psi}_1=
(-1)^{\ket{\Psi}} \brai{12}{V_2}\ket{\Psi}_1\ket{V_2}_{32}=
(-1)^{\ket{\Psi}} \ket{\Psi}_2^\bpz \ket{V_2}_{32},
\end{equation}
where as usual by $(-1)^{\ket{\Psi}}$ we mean $(-1)$ to the power of
the parity of ${\ket{\Psi}}$ and $\bra{\Psi}=\ket{\Psi}^\bpz$ stands
for the BPZ conjugate of $\ket{\Psi}$. Now, we use
$(-1)^{\ket{\Psi}}=-(-1)^{\bra{\Psi}}$ to get,
\begin{equation}
\braketi{2}{\Psi}{V_2}_{23}=-\braketi{2}{\Psi}{V_2}_{32}=
(-1)^{\bra{\Psi}} \brai{2}{\Psi}^\bpz\,.
\end{equation}
It is also possible to define a vertex that implements BPZ
conjugation. This vertex is symmetric and differs from $\ket{V_2}$
only by zero modes,
\begin{equation}
\ket{\tilde V_2}_{12}=
 -\exp \!\bigg(\!-\sum_{n=2}^\infty \Big
  (b^{1}_{-n} (-1)^n c^{2}_{-n}+b^{2}_{-n} (-1)^n c^{1}_{-n}
 \Big)\bigg)
 (c_1^1+c_{-1}^2)(c_0^1-c_0^2)(c_{-1}^1+c_1^2)\ket{0}_{12}.
\end{equation}

The ghost three-vertex is constructed over
the triple tensor product of the vacuum
$\bra{\vb}$~\cite{Gross:1987fk},
\begin{equation}
\label{GJver}
\brai{321}{V_3}=N_3\,\brai{321}{\vb}
  \exp\Big(\sum_{r,s=1}^3\sum_{m=0}^\infty\sum_{n=1}^\infty
  b^{r}_{m} V^{rs}_{mn}c^{s}_{n}\Big)\,.
\end{equation}
The normalization constant is,
\begin{equation}
N_3 =K^3\,,\qquad K=\frac{3\sqrt{3}}{4}\,,
\end{equation}
and the coefficients $V^{rs}_{mn}$
are~\cite{LeClair:1989sp,Taylor:2003gn},
\begin{equation}
V^{rs}_{mn}=\oint \frac{dz\, dw}{(2\pi i)^2}\frac{1}{z^{n-1}w^{m+2}}
  \frac{1}{f_s(z)-f_r(w)}\frac{f_s'(z)^2}{f_r'(w)}
   \prod_{i=1}^3 \frac{f_r(w)-f_i(0)}{f_s(z)-f_i(0)}\,.
\end{equation}
The functions $f_r$ are the conformal transformations defining
the vertex,
\begin{equation}
f_r(z)=e^{-\frac{2\pi i}{3}r}
 \Big(\frac{1+i z}{1-i z}\Big)^\frac{2}{3}\,.
\end{equation}
The coefficients can be efficiently calculated by
the method in~\cite{Rastelli:2001jb,Taylor:2003gn}
for the fermionic ghost. We define,
\begin{equation}
\Big(\frac{1+ix}{1-ix}\Big)^{\frac{1}{3}}=a_0+ia_1+a_2+ia_3+\dots\,,
  \qquad
\Big(\frac{1+ix}{1-ix}\Big)^{\frac{2}{3}}=b_0+ib_1+b_2+ib_3+\dots\,.
\end{equation}
In terms of which $V^{rs}_{mn}$ are given by,
\begin{equation}
\begin{aligned}
V^{rr}_{mn}&=\left\{\begin{array}{ll}
\frac{1}{3}(a_n^2+2(-1)^n a_n b_n+(-1)^n-2
     \sum _{k=0}^n (-1)^{n-k} a_k^2)\\
 \frac{2 (-1)^m n }{3(m^2-n^2)}(m a_n b_m-n a_m b_n)\\
 0   \end{array}\right.
 &\begin{array}{ll} m=n\\ m \neq n,\ m\equiv_2 n\\
      m-n\equiv_2 1 \end{array}  \\
V^{r(r\pm 1)}_{mn}&=\left\{\begin{array}{ll}
\frac{1}{2}\big( (-1)^n-V^{rr}_{nn} \big) \\
 -\frac{V^{rr}_{mn} }{2} \\
 - \frac{n (\mp 1)^{m+n}}{\sqrt{3} (m^2-n^2)}(m a_n b_m+n a_m b_n)
\end{array}\right.
 &\begin{array}{ll} m=n\\ m \neq n,\ m\equiv_2 n\\
     m-n\equiv_2 1 \end{array}
\end{aligned}
\end{equation}

The coefficients $V^{rs}_{mn}$ are symmetric with respect to cyclic
permutation of the indices $r,s$. The vacuum $\brai{321}{\vb}$ is
antisymmetric with respect to permutation of any two indices
and is, therefore, symmetric with respect to cyclic
permutation. Thus, $\bra{V_3}$ is symmetric with respect to cyclic
permutation. The same should be true for any vertex, due to the
geometric picture of gluing. It may seem that the parity of the states
on which the vertex operate can modify this symmetry
property. However, since $\bra{V_N}$ is odd ($\forall N$), there
should be an odd number of odd states on which the vertex operates
in order to get a non-zero result. A cyclic permutation among the $N$
states induces a cyclic permutation among the odd ones. A cyclic
permutation of an odd number of objects is even. Thus, the symmetry
property holds as stated.

\subsection{The anomaly}

We know from~(\ref{VnIdentities}) that
\begin{equation}
\label{V3V1V2}
\braketi{321}{V_3}{V_1}_1=\brai{32}{V_2}.
\end{equation}
We want to examine this equality in level truncation.

In order to evaluate this overlap, we have to generalize eq.~(19)
of~\cite{Kostelecky:2000hz} to the case of linear terms in both
sides. The generalization is straightforward and reads,
\begin{equation}
\begin{aligned}
\bra{\Om}&\exp(-b S c+b\la^c+\la^b c)
   \exp(b^\her V c^\her+b^\her\mu^c+\mu^b c^\her)\ket{\Om}\\&=
\det(\One-S V)\exp\big(\mu^b(\One-S V)^{-1}(S\mu^c-\la^c)+
  \la^b(\One-V S)^{-1}(\mu^c-V\la^c)\big)\,.
\end{aligned}
\end{equation}
Here, $\la^{b,c}$ and $\mu^{b,c}$ are given anticommuting vectors,
$S,V$ are given matrices and the summation is everywhere implicit.
The vacua $\bra{\Om}$, $\ket{\Om}$ are canonically conjugate
such that the $b^\her,c^\her$ are creation operators and the $b,c$ are
destruction operators. It would be
convenient here to use $\bra{\vb}$ and $\ket{\va}$ for the
vacua. Thus, $b_0$ would be considered as a destruction operator,
while $c_0^\her=c_0$ would be a creation operator.
Actually, due to the form of~(\ref{V1}), what we really need is,
\begin{equation}
\begin{aligned}
\bra{\vb}&\exp(-b S c+b\la^c+\la^b c)(b^\her v)
   \exp(b^\her V c^\her)\ket{\va}\\&=(-1)^{\ket{\va}}
\overleftarrow{\partial}\!_{\e}
   \Big(\det(\One-S V)
     \exp\big(\la^b(\One-V S)^{-1}(v \e-V\la^c)\big)\Big)\\&=
  -\det(\One-S V)\big(\la^b(\One-V S)^{-1}v\big)
    \exp\big(-\la^b(\One-V S)^{-1}V\la^c)\big)\,.
\end{aligned}
\end{equation}
Here, $v$ is a given commuting vector, $\e$ is an anticommuting
parameter, and
$\overleftarrow{\partial}\!_{\e}$
is a derivative from the right with
respect to $\e$.

We now use~(\ref{V1},\ref{GJver}) to write the ghost part
of~(\ref{V3V1V2}) as,
\begin{equation}
\begin{aligned}
\braketi{321}{V_3}{V_1}_1=&N_3\,\brai{32}{\vb}
  \exp\Big(\sum_{r,s=2}^3 b^r V^{rs} c^s\Big)\cdot\\&\cdot
\brai{1}{\vb}\exp\big(-b^1 S c^1+b^1\la^c+\la^b c^1\big)
 (b^{1\her} v)\exp\big(b^{1\her} V c^{1\her}\big) \ket{\va}_1,
\end{aligned}
\end{equation}
where we define,
\begin{equation}
\begin{aligned}
&S=-V^{11}\,,\qquad \la^c=\sum_{s=2}^3 V^{1s}c^s\,,\qquad
 \la^b=\sum_{r=2}^3 b^r V^{r1}\,,\\
&v_m=\left\{\begin{array}{ll}
   0 &\quad m\equiv_2 0 \\(-1)^{\frac{m-1}{2}} &\quad m\equiv_2 1
  \end{array} \right.\,,\qquad
 V_{mn}=\left\{\begin{array}{ll}
   2(-1)^{\frac{m}{2}} &\quad n=0,\ m\equiv_2 0\\
   (-1)^{m+1} &\quad m=n\\
   0 &\quad \text{otherwise}
    \end{array} \right.\,.
\end{aligned}
\end{equation}

We would like to compare the result of this calculation
with~(\ref{V2up}).
Define $\sigma^r_m$ and $U^{rs}_{mn}$ by,
\begin{equation}
\sigma^r=V^{r1}(\One-V S)^{-1}v\,,\qquad
U^{rs}=V^{rs} - V^{r1}(\One-V S)^{-1}V V^{1s}\,.
\end{equation}
It is seen that in order that~(\ref{V3V1V2}) would hold,
$\la^b(\One-V S)^{-1}v$ should be proportional to $b_0^2-b_0^3$.
That is, as we increase the level we expect,
\begin{equation}
\begin{aligned}
&\frac{\sigma^2_0}{\sigma^3_0}\rightarrow -1\,,\\
&\frac{\sigma^r_m}{\sigma^3_0}\rightarrow 0 \qquad
   \forall m \neq 0\,,\, \forall r\,.
\end{aligned}
\end{equation}
In fact, the symmetry of the equations implies that the first of these
equations exactly holds for any level. We check the maximal absolute
value of the second expression in level truncation,
from level 40 to 400, which we then fit to the form,
\begin{equation}
\label{rDef}
r_\ell
\equiv \max_{m}\left|\frac{\sigma^3_m}{\sigma^3_0}\right|=
  a+\frac{b}{\ell^c} \,. 
\end{equation}
The results are summarized in figure~\ref{fig:bFit}.
\FIGURE{
\centerline{ \epsfig{figure=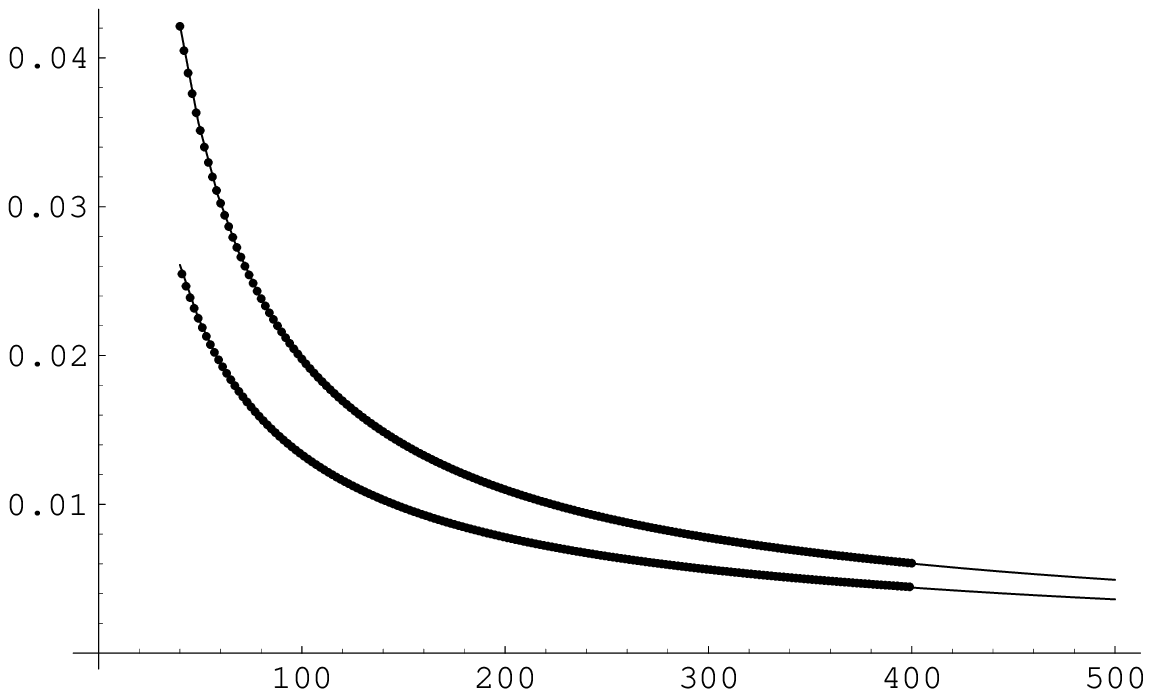,width=9.5cm} }
\caption{A fit for the ratio $r_\ell$ for even (top curve) and odd
  levels. For the even levels we get $a=-0.00057, b=0.86, c=0.81$. For
  the odd levels we get $a=-0.0011, b=0.35, c=0.69$. The data seems to
  indicate that this ratio vanishes as it should.
}
\label{fig:bFit}
}

Next, we have to check that in the $\ell\rightarrow\infty$
limit, $U^{22}_{mn},U^{33}_{mn}\rightarrow 0$,
while $U^{23},U^{32}\rightarrow C$, where now,
\begin{equation}
C_{mn}=(-1)^n\delta_{m,n}\,,\qquad m\geq 0\,,\,n\geq 1\,.
\end{equation}
First, we consider the $m=0$ rows. These terms need not vanish.
Such terms would not contribute, provided two restrictions hold:
$U^{rs}_{0n}r_\ell \rightarrow 0$,
in order to cancel terms of the type $(b_0^2-b_0^3) b_m^r c^s_n$
and $U^{22}_{0n}+U^{23}_{0n}\rightarrow 0$, in order to cancel
factors that would take the form
$(b^2_0 b^3_0+b^3_0 b^2_0)c^s_k=0$. We can calculate,
\begin{equation}
\begin{aligned}
U^{22}+U^{23}=&V^{22}+V^{23}-V^{21}
  (\One+V V^{11})^{-1}V(V^{12}+V^{13})=\\&
  V^{22}+V^{23}-V^{21}(\One+V V^{11})^{-1}V(C-V^{11})=\\&
  V^{22}+V^{23}-V^{21}(\One+V V^{11})^{-1}(-\One-V V^{11})=
  V^{22}+V^{23}+V^{21}=C\,,
\label{UU}
\end{aligned}
\end{equation}
and all the manipulations performed are valid at any finite level.
Thus, the second restriction automatically holds.
The first restriction is checked numerically.
For a given level we discard the last 20 rows and columns of
this matrix. We see that from level $140$ to level $400$,
$\max|U^{22}_{0n}|$ increases from $1.19$ to $1.34$.
Combined with the data for~(\ref{rDef}),
it seems that the last limit also holds well.
With the help of~(\ref{UU}), in order to ensure that the exponent
behaves correctly, it is enough to check that
\begin{equation}
U^{22}_{mn}\rightarrow 0\,\quad m>0\,.
\end{equation}
For $\max(|U^{22}_{mn}|)$ we used a fit of the form,
\begin{equation}
\label{rDef2}
\hat r_\ell
\equiv \max_{m,n\geq 1}\left|U^{22}_{mn}\right|=
  a+\frac{b}{\ell^c}\,.
\end{equation}
These results are summarized in figure~\ref{fig:ExFit}.
\FIGURE{
\centerline{ \epsfig{figure=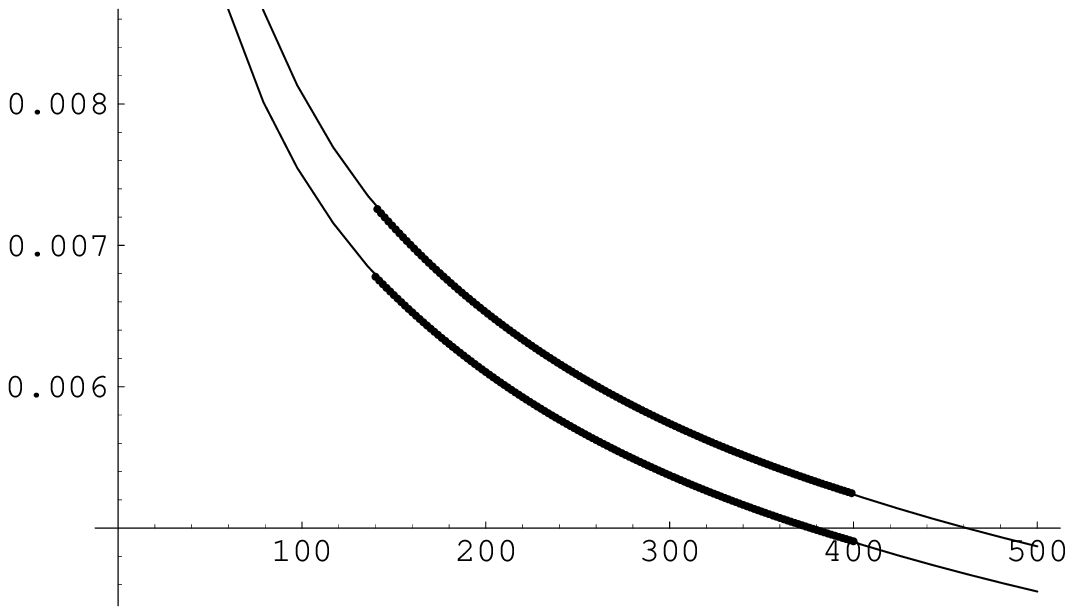,width=9.5cm} }
\caption{A fit of $\max|U^{22}_{mn}|$ for even (bottom curve) and odd
  levels. For even levels we get $a=-0.0026, b=0.027, c=0.21$. For odd
  levels we get $a=-0.0012, b=0.031, c=0.26$. Again, the data seems
  consistent with the vanishing of $U_{22}$.
}
\label{fig:ExFit}
}

What is left now is to determine the normalization,
\begin{equation}
\gamma_{13}=\gamma_{13}^{\text{gh}}\gamma_{13}^{\text{m}}\,.
\end{equation}
The ghost
part of the normalization is,
\begin{equation}
\gamma_{13}^{\text{gh}}= N_3
\sigma_0^2 \det(\One-S V)
  \rightarrow \infty\,.
\end{equation}
This factor should be multiplied by the matter factor,
\begin{equation}
\gamma_{13}^{\text{m}}=\det(\One-V^{11\text{ m}} C)^{-\frac{26}{2}}
\rightarrow 0\,.
\end{equation}
Here, the matter three-vertex $V^{11\text{ m}}$ and $C$ do not
have zero modes.
The factor of $26$ comes from the critical dimension and indeed
$\gamma_{13}$
diverges logarithmically in any other dimension.
The $\gamma_{13}$ fit is summarized in figure~\ref{fig:DetFit}.
\FIGURE{
\centerline{ \epsfig{figure=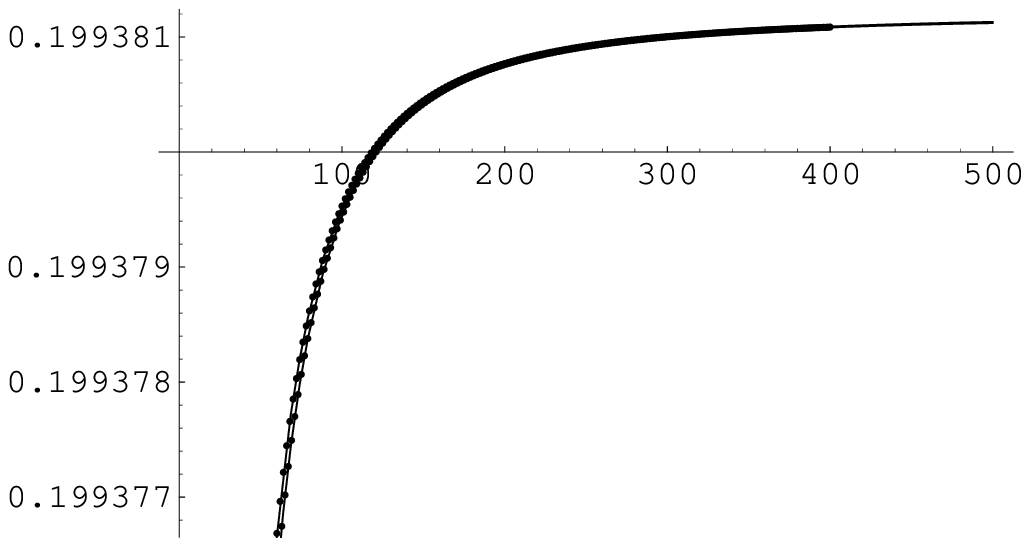,width=9.5cm} }
\caption{A fit of the normalization factor
  $\gamma_{13}=a+b/\ell^c$. Odd and even levels behave almost the same
  in this case, $b=-0.0124,-0.0196$ and $c=1.93,2.03$ for even and odd
  levels respectively. The limit value in both cases is $a=0.1994$,
  which is again anomalous.}
\label{fig:DetFit}
}

It is possible to reinstate the correct limit value for $\gamma_{13}$
by using different levels for the matter and ghost
parts. For a linear relation between $\ell^{\m}$ and $\ell^{\g}$,
the normalization $N_2$ would converge and for,
\begin{equation}
\ell^{\g}\approx 3.07 \ell^{\m}\,,
\end{equation}
the limit value is unity. However, it is not clear what is the meaning
of such a relation or whether it is a universal value for level
truncation calculations in the discrete fermionic basis. We believe
that it is not universal and meaningless and that the discrepancy
shows that level truncation is not adequate for such calculations.

\section{Field level truncation}
\label{fieldLT}

In this section we show how anomalies can emerge in field level
truncation calculations, as well as in formal calculations using the
Virasoro algebra. In particular, we claim that the naive unity
normalization of string vertices and of wedge states should be
reconsidered~\cite{LeClair:1989sp}.

For the field level truncation it is simpler to compute the overlap of
the identity state with the $n=3$ wedge state over the vacuum
$\bra{3}$ ($\bra{3}$ here is the vacuum state with ghost number
three~(\ref{The3Vac})
and should not be confused with the fact that $\bra{n}$ is sometimes
used for wedge
states).
This vacuum choice corresponds to choosing
$c(0)c'(0)c''(0)$ as the operator that saturates the ghost number.
This computation is
equivalent to the descent relation calculation of~(\ref{DescRel}),
as can be seen by contracting~(\ref{DescRel}) with
$\bra{0}\bra{3}$.

Wedge states are surface states and as such can be generated
purely from Virasoro generators.
In field level truncation, Virasoro generators are very convenient
to work with, since their index is their level.
This makes the calculation trivial.
Because the central charge is zero, we can use the relation,
\begin{equation}
\label{LevelTrun}
\bra{3}f(L_2,\ldots,L_{n})g(L_{-2},\ldots,L_{-m})\ket{0}=
f(0,\dots,0)g(0,\dots,0)\,.
\end{equation}
In our case we have surface states, which are exponents of a linear
combination of Virasoro generators, so $f(0,..,0)g(0,..,0)=1$.
This relation holds for any finite level.

There can be subtleties in this
relation, since it does not guaranty that the states exist at all.
It is impossible to notice a non-legitimate state by working in
a finite level with central charge $c=0$.
This problem raises a question on the validity of level truncation.
Consider the regularized butterfly state as a simple test case,
\begin{equation}
\label{regBut}
\ket{\B_a}=\exp(-\frac{a}{2} L_{-2})\ket{0}\,.
\end{equation}
The states with $a>1$ are not legitimate since they do not have a well
defined local coordinate
patch~\cite{Gaiotto:2001ji,Gaiotto:2002kf,Fuchs:2004xj}.
However, the overlap of such a state with any other surface state
is unity at any level, with no sign of the singularity at $\ell=\infty$.

We propose a criterion to distinguish when eq.~(\ref{LevelTrun})
can be trusted. It
is based on the observation that the singularity can be noticed
when we separate the matter and ghost sectors.
The procedure is to calculate the overlap for
the case of a non-zero central charge. We call this quantity $N(c)$.
It obeys the relation,
\begin{equation}
\label{NRel}
N(c_1)N(c_2)=N(c_1+c_2)\,.
\end{equation}
This can be seen by taking a Virasoro algebra with central charge
$c_1+c_2$, which is the sum of two independent Virasoro algebras with
central charges $c_1, c_2$.
Repeated use of this identity and continuity with respect to $c$ imply
that
\begin{equation}
N(c)=\al^c\,,
\end{equation}
where $\al$ depends on the states being evaluated.
Our criterion is that $N(c)$ should be continuous and regular,
meaning $\al$ should be in the range,
\begin{equation}
0<\al<\infty\,.
\end{equation}
Only when this criterion is met, can we trust the result $N(0)=1$.

Moreover, we require that the norm of each state should meet
this criterion.
States with $\al<0$ are not legitimate and should be discarded.
At most they can be treated formally as in~\cite{Bars:2002nu}.
States with $\al=0,\infty$ belong to the boundary of the space of
legitimate states.
Such states have zero norm in the ghost sector and infinite norm in
the matter sector or vice versa.
A familiar example for such a state is the
sliver~\cite{Moore:2001fg}.
It might be possible to regularize such
a state by considering a one-parameter family of regular states
leading to it.
However, anomalous normalization factors should be expected in this
case.
In the case of the regulated butterfly~(\ref{regBut}), a
straightforward calculation gives (assuming that vacuum ghost number
was properly chosen),
\begin{equation}
N_{a,b}(c)\equiv \braket{\B_a}{\B_b}=\big(1-(ab)^2\big)^{-c/8}\,.
\end{equation}
Here we see that indeed the states with $a>1$ are not legitimate,
while the canonical butterfly is marginal.
We also calculated $N_{a,a}(c)$ in
oscillator level truncation for a single matter
field, $c=1$. We got very good convergence to the expected
results for $|a|<1$.

For other surface states, we cannot
calculate $N(c)$ so easily
and in the general case the calculation seems impossible.
We can calculate the general case in field level truncation, but
a calculation for finite $\ell$ will never give us a conclusive
result on the $\ell\rightarrow\infty$ limit.
Alternatively, we can use oscillator level truncation in the matter
sector.

For the wedge states we have a third option. We can use the continuous
basis, which gives the oscillatory level truncated result at infinite
level (with some subtleties regarding
the way margins should be handled,
as mentioned in section~\ref{sec:bosonic}).
The normalization factor is an integral involving the spectral
density. Since the leading term in the spectral density is a divergent
constant~(\ref{specDens}), states with a non-zero integral have
divergent (or zero) normalization in the matter sector. The only
surface states in $\HH_{\ka^2}$ are the butterflies and the
wedges~\cite{Fuchs:2004xj,Uhlmann:2004mv}. All these states
(other than the vacuum) have a non-zero integral.
Thus, they are marginal. Indeed, it is known that
the non-balanced wedge states are anomalous~\cite{Schnabl:2002gg}.
It is also known that rank one projectors (as are the butterflies) are
marginal. In this case the singularity comes from one eigenvector with
the eigenvalue 1~\cite{Gaiotto:2002kf}\footnote{In~\cite{Fuchs:2002zz}
  it was claimed that some non-surface-state rank-one projectors can
  have a $-1$ eigenvalue instead. However, since such projectors
  approach the identity string field as $\ka\rightarrow 0$, they
  probably should not be considered as rank-one projectors. In fact,
  it can be seen that the transformation that such states induce to a
  half-string basis~\cite{Fuchs:2003wu} becomes singular in this
  limit, so their left-right factorization is ill-defined.}, as
opposed to the continuous case where the singularity comes from the
density of the eigenvalues.

\section{Conclusions}
\label{sec:conc}

In this paper we studied various level truncation schemes.
While it seems that the correct results are obtained
for calculations involving regular states, we found anomalous
normalization factors for singular states.
These singular states include rank-one projectors and wedge
states\footnote{Rank-one projectors are important since they
are the solutions of vacuum string field theory.
It is possible to claim that they are singular due to the
singular nature of vacuum string field theory.
However, there are quite a few interesting results that were
achieved within vacuum string field theory using projectors.
Wedge states probably have to be
included in the yet unknown space of string fields.
For the BRST construction of string field theory, states of
arbitrary ghost number should be considered. Due to the central
role played by the star product in string field theory, it is
natural to expect that this space would form a star-algebra.
The $n=1$ wedge state is $V_1$, which we must consider, at
least as a functional, while all wedge states with integer $n>1$
are generated in the star-algebra from the $n=2$ wedge state,
i.e. the vacuum $\ket{0}$.}, as well as the string
vertices that define the star product. Thus, these states cannot be
ignored.
We also illustrated the relation between level truncation results and
continuous basis evaluation. Apparently, continuous basis results are
obtained from level truncation by using margins for the evaluation of
non-linear expressions and then sending the margin factor to infinity.
These results should be taken into account whenever using level
truncation with potentially singular states.

The most dramatic consequence of our findings is related to the issue
of normalization factors of surface states and string vertices.
It is always assumed that such states have a unity normalization
factor. The reason for this is the discussion around
eq.~(2.39) of~\cite{LeClair:1989sj}, where a $c=0$ Virasoro algebra is
considered. There, it is claimed that given two sets of coefficients
$v_{-n}^{(i)},\ n=2..\infty,\,i=1,2$,
the following equality holds for some $v_{-n}^{(i)}(t,s),\ i=3,4$,
\begin{equation}
\label{LPPeq}
\exp\Big(t \sum_{n=2}^\infty v^{(1)}_{-n}L_n\Big)
  \exp\Big(s \sum_{n=2}^\infty v^{(2)}_{-n}L_{-n}\Big)=
\exp\Big(\sum_{n=2}^\infty v^{(3)}_{-n}L_{-n}\Big)
  \exp\Big(\sum_{n=-1}^\infty v^{(4)}_{-n}L_n\Big)\,.
\end{equation}
From this equality the unity normalization follows after contracting
with the left and right vacua.
However, this assertion is based on a theorem that can be applied only
for small enough $s,t$.
This assumption is unjustified for singular states.

The singularity of the vertices could lead one to believe that all
string field theory calculations would suffer from this anomaly.
This is not the case since only the two-vertex and three-vertex
appear in the action. The former
is anomaly free, while the anomaly of the later can be absorbed by
a redefinition of the string coupling constant~\cite{Belov:2003qt}.
It should be interesting to study the impact of these effects for
non-polynomial string field theories.

All this may support Belov's suggestion that the string vertices
should be prefactored by specific partition
functions~\cite{Belov:2003qt}. Belov evaluated these factors using
consistency requirements in the continuous basis. However, in light of
the discussion above, we cannot trust these results either, due to the
implicit use of level truncation in continuous basis regularizations.
We cannot even be sure that these factors are universal for continuous
basis calculations.

Thus, we find that it is desirable to evaluate these factors
directly. Another important direction would be to find a
regularization that respects all the symmetries, as well as the
descent relations, in order to calculate the normalization of the
vertices directly. These tasks would teach us more on the range of
validity of level truncation, on the limitations of using
eq.~(\ref{LPPeq}) and would be important for concrete future
calculations in string field theory.

\section*{Acknowledgments}

We would like to thank
Ofer Aharony, Dmitriy Belov and Alon Marcus, for discussions.
We would like to thank Roman Gorbachev and Dmitry Rychkov for bringing
a typo to our attention.
E.~F. would like to thank Tel-Aviv
University, where part of this work was performed, for hospitality.
M.~K. would like to thank the Albert Einstein
Institut, where part of this work was performed, for hospitality.
This work was supported in part by the German-Israeli Foundation
for Scientific Research and by the Israel Science Foundation.


\bibliography{FK}

\providecommand{\href}[2]{#2}\begingroup\raggedright\begin{thebibliography}{10}

\bibitem{Witten:1986cc}
E.~Witten, {\it Noncommutative geometry and string field theory},  {\em Nucl.
  Phys.} {\bf B268} (1986) 253.

\bibitem{Kostelecky:1990nt}
V.~A. Kostelecky and S.~Samuel, {\it On a nonperturbative vacuum for the open
  bosonic string},  {\em Nucl. Phys.} {\bf B336} (1990) 263.

\bibitem{Sen:1999xm}
A.~Sen, {\it Universality of the tachyon potential},  {\em JHEP} {\bf 12}
  (1999) 027, [\href{http://xxx.lanl.gov/abs/hep-th/9911116}{{\tt
  hep-th/9911116}}].

\bibitem{Sen:1999nx}
A.~Sen and B.~Zwiebach, {\it Tachyon condensation in string field theory},
  {\em JHEP} {\bf 03} (2000) 002,
  [\href{http://xxx.lanl.gov/abs/hep-th/9912249}{{\tt hep-th/9912249}}].

\bibitem{Rastelli:2000iu}
L.~Rastelli and B.~Zwiebach, {\it Tachyon potentials, star products and
  universality},  {\em JHEP} {\bf 09} (2001) 038,
  [\href{http://xxx.lanl.gov/abs/hep-th/0006240}{{\tt hep-th/0006240}}].

\bibitem{Gaiotto:2002wy}
D.~Gaiotto and L.~Rastelli, {\it Experimental string field theory},  {\em JHEP}
  {\bf 08} (2003) 048, [\href{http://xxx.lanl.gov/abs/hep-th/0211012}{{\tt
  hep-th/0211012}}].

\bibitem{Kostelecky:1995qk}
V.~A. Kostelecky and R.~Potting, {\it Expectation values, {L}orentz invariance,
  and {CPT} in the open bosonic string},  {\em Phys. Lett.} {\bf B381} (1996)
  89--96, [\href{http://xxx.lanl.gov/abs/hep-th/9605088}{{\tt
  hep-th/9605088}}].

\bibitem{Rastelli:2001jb}
L.~Rastelli, A.~Sen, and B.~Zwiebach, {\it Classical solutions in string field
  theory around the tachyon vacuum},  {\em Adv. Theor. Math. Phys.} {\bf 5}
  (2002) 393--428, [\href{http://xxx.lanl.gov/abs/hep-th/0102112}{{\tt
  hep-th/0102112}}].

\bibitem{Taylor:2002bq}
W.~Taylor, {\it Perturbative diagrams in string field theory},
  \href{http://xxx.lanl.gov/abs/hep-th/0207132}{{\tt hep-th/0207132}}.

\bibitem{Potting:1988ra}
R.~Potting and C.~Taylor, {\it The midpoint transformation in {W}itten's string
  field theory},  {\em Nucl. Phys.} {\bf B316} (1989) 59.

\bibitem{Okuyama:2002tw}
K.~Okuyama, {\it Ratio of tensions from vacuum string field theory},  {\em
  JHEP} {\bf 03} (2002) 050,
  [\href{http://xxx.lanl.gov/abs/hep-th/0201136}{{\tt hep-th/0201136}}].

\bibitem{Bars:2002bt}
I.~Bars and Y.~Matsuo, {\it Associativity anomaly in string field theory},
  {\em Phys. Rev.} {\bf D65} (2002) 126006,
  [\href{http://xxx.lanl.gov/abs/hep-th/0202030}{{\tt hep-th/0202030}}].

\bibitem{Rastelli:2001hh}
L.~Rastelli, A.~Sen, and B.~Zwiebach, {\it Star algebra spectroscopy},  {\em
  JHEP} {\bf 03} (2002) 029,
  [\href{http://xxx.lanl.gov/abs/hep-th/0111281}{{\tt hep-th/0111281}}].

\bibitem{Feng:2002rm}
B.~Feng, Y.-H. He, and N.~Moeller, {\it The spectrum of the {N}eumann matrix
  with zero modes},  {\em JHEP} {\bf 04} (2002) 038,
  [\href{http://xxx.lanl.gov/abs/hep-th/0202176}{{\tt hep-th/0202176}}].

\bibitem{Belov:2002fp}
D.~M. Belov, {\it Diagonal representation of open string star and {M}oyal
  product},  \href{http://xxx.lanl.gov/abs/hep-th/0204164}{{\tt
  hep-th/0204164}}.

\bibitem{Erler:2002nr}
T.~G. Erler, {\it Moyal formulation of {W}itten's star product in the fermionic
  ghost sector},  \href{http://xxx.lanl.gov/abs/hep-th/0205107}{{\tt
  hep-th/0205107}}.

\bibitem{Belov:2002pd}
D.~M. Belov and A.~Konechny, {\it On continuous {M}oyal product structure in
  string field theory},  {\em JHEP} {\bf 10} (2002) 049,
  [\href{http://xxx.lanl.gov/abs/hep-th/0207174}{{\tt hep-th/0207174}}].

\bibitem{Fuchs:2002zz}
E.~Fuchs, M.~Kroyter, and A.~Marcus, {\it Squeezed state projectors in string
  field theory},  {\em JHEP} {\bf 09} (2002) 022,
  [\href{http://xxx.lanl.gov/abs/hep-th/0207001}{{\tt hep-th/0207001}}].

\bibitem{Ihl:2003fw}
M.~Ihl, A.~Kling, and S.~Uhlmann, {\it String field theory projectors for
  fermions of integral weight},  {\em JHEP} {\bf 03} (2004) 002,
  [\href{http://xxx.lanl.gov/abs/hep-th/0312314}{{\tt hep-th/0312314}}].

\bibitem{Okuyama:2002yr}
K.~Okuyama, {\it Ghost kinetic operator of vacuum string field theory},  {\em
  JHEP} {\bf 01} (2002) 027,
  [\href{http://xxx.lanl.gov/abs/hep-th/0201015}{{\tt hep-th/0201015}}].

\bibitem{Fuchs:2002wk}
E.~Fuchs, M.~Kroyter, and A.~Marcus, {\it Virasoro operators in the continuous
  basis of string field theory},  {\em JHEP} {\bf 11} (2002) 046,
  [\href{http://xxx.lanl.gov/abs/hep-th/0210155}{{\tt hep-th/0210155}}].

\bibitem{Belov:2002sq}
D.~M. Belov and A.~Konechny, {\it On spectral density of {N}eumann matrices},
  {\em Phys. Lett.} {\bf B558} (2003) 111--118,
  [\href{http://xxx.lanl.gov/abs/hep-th/0210169}{{\tt hep-th/0210169}}].

\bibitem{Belov:2003qt}
D.~M. Belov, {\it Witten's ghost vertex made simple (bc and bosonized ghosts)},
   {\em Phys. Rev.} {\bf D69} (2004) 126001,
  [\href{http://xxx.lanl.gov/abs/hep-th/0308147}{{\tt hep-th/0308147}}].

\bibitem{Beccaria:2003ak}
M.~Beccaria and C.~Rampino, {\it Level truncation and the quartic tachyon
  coupling},  {\em JHEP} {\bf 10} (2003) 047,
  [\href{http://xxx.lanl.gov/abs/hep-th/0308059}{{\tt hep-th/0308059}}].

\bibitem{LeClair:1989sp}
A.~LeClair, M.~E. Peskin, and C.~R. Preitschopf, {\it String field theory on
  the conformal plane. 1. {K}inematical principles},  {\em Nucl. Phys.} {\bf
  B317} (1989) 411.

\bibitem{LeClair:1989sj}
A.~LeClair, M.~E. Peskin, and C.~R. Preitschopf, {\it String field theory on
  the conformal plane. 2. {G}eneralized gluing},  {\em Nucl. Phys.} {\bf B317}
  (1989) 464.

\bibitem{Fuchs:2003wu}
E.~Fuchs, M.~Kroyter, and A.~Marcus, {\it Continuous half-string representation
  of string field theory},  {\em JHEP} {\bf 11} (2003) 039,
  [\href{http://xxx.lanl.gov/abs/hep-th/0307148}{{\tt hep-th/0307148}}].

\bibitem{Kostelecky:2000hz}
V.~A. Kostelecky and R.~Potting, {\it Analytical construction of a
  nonperturbative vacuum for the open bosonic string},  {\em Phys. Rev.} {\bf
  D63} (2001) 046007, [\href{http://xxx.lanl.gov/abs/hep-th/0008252}{{\tt
  hep-th/0008252}}].

\bibitem{Polchinski:1998rq}
J.~Polchinski, {\it String theory. vol. 1: An introduction to the bosonic
  string}, . Cambridge, UK: Univ. Pr. (1998) 402 p.

\bibitem{Zwiebach:1997fe}
B.~Zwiebach, {\it Oriented open-closed string theory revisited},  {\em Annals
  Phys.} {\bf 267} (1998) 193--248,
  [\href{http://xxx.lanl.gov/abs/hep-th/9705241}{{\tt hep-th/9705241}}].

\bibitem{Maccaferri:2003rz}
C.~Maccaferri and D.~Mamone, {\it Star democracy in open string field theory},
  {\em JHEP} {\bf 09} (2003) 049,
  [\href{http://xxx.lanl.gov/abs/hep-th/0306252}{{\tt hep-th/0306252}}].

\bibitem{Fuchs:2004xj}
E.~Fuchs and M.~Kroyter, {\it On surface states and star-subalgebras in string
  field theory},  {\em JHEP} {\bf 10} (2004) 004,
  [\href{http://xxx.lanl.gov/abs/hep-th/0409020}{{\tt hep-th/0409020}}].

\bibitem{Gross:1987fk}
D.~J. Gross and A.~Jevicki, {\it Operator formulation of interacting string
  field theory. 2},  {\em Nucl. Phys.} {\bf B287} (1987) 225.

\bibitem{Bonora:2003xp}
L.~Bonora, C.~Maccaferri, D.~Mamone, and M.~Salizzoni, {\it Topics in string
  field theory},  \href{http://xxx.lanl.gov/abs/hep-th/0304270}{{\tt
  hep-th/0304270}}.

\bibitem{Kling:2003sb}
A.~Kling and S.~Uhlmann, {\it String field theory vertices for fermions of
  integral weight},  {\em JHEP} {\bf 07} (2003) 061,
  [\href{http://xxx.lanl.gov/abs/hep-th/0306254}{{\tt hep-th/0306254}}].

\bibitem{Taylor:2003gn}
W.~Taylor and B.~Zwiebach, {\it D-branes, tachyons, and string field theory},
  \href{http://xxx.lanl.gov/abs/hep-th/0311017}{{\tt hep-th/0311017}}.

\bibitem{Gaiotto:2001ji}
D.~Gaiotto, L.~Rastelli, A.~Sen, and B.~Zwiebach, {\it Ghost structure and
  closed strings in vacuum string field theory},
  \href{http://xxx.lanl.gov/abs/hep-th/0111129}{{\tt hep-th/0111129}}.

\bibitem{Gaiotto:2002kf}
D.~Gaiotto, L.~Rastelli, A.~Sen, and B.~Zwiebach, {\it Star algebra
  projectors},  {\em JHEP} {\bf 04} (2002) 060,
  [\href{http://xxx.lanl.gov/abs/hep-th/0202151}{{\tt hep-th/0202151}}].

\bibitem{Bars:2002nu}
I.~Bars and Y.~Matsuo, {\it Computing in string field theory using the {M}oyal
  star product},  {\em Phys. Rev.} {\bf D66} (2002) 066003,
  [\href{http://xxx.lanl.gov/abs/hep-th/0204260}{{\tt hep-th/0204260}}].

\bibitem{Moore:2001fg}
G.~Moore and W.~Taylor, {\it The singular geometry of the sliver},  {\em JHEP}
  {\bf 01} (2002) 004, [\href{http://xxx.lanl.gov/abs/hep-th/0111069}{{\tt
  hep-th/0111069}}].

\bibitem{Uhlmann:2004mv}
S.~Uhlmann, {\it A note on kappa-diagonal surface states},  {\em JHEP} {\bf 11}
  (2004) 003, [\href{http://xxx.lanl.gov/abs/hep-th/0408245}{{\tt
  hep-th/0408245}}].

\bibitem{Schnabl:2002gg}
M.~Schnabl, {\it Wedge states in string field theory},  {\em JHEP} {\bf 01}
  (2003) 004, [\href{http://xxx.lanl.gov/abs/hep-th/0201095}{{\tt
  hep-th/0201095}}].

\end{thebibliography}\endgroup

\end{document}